# Performance Advantages of Monolithically Patterned Wide-Narrow-Wide All-Graphene on Insulator Devices

Dincer Unluer, Frank Tseng, Avik W. Ghosh, and Mircea R. Stan

**Abstract**—We investigate theoretically the performance advantages of all-graphene nanoribbon field-effect transistors (GNRFETs) whose channel and source/drain (contact) regions are patterned *monolithically* from a single sheet of graphene. In our simulated devices, the source/drain and interconnect regions are composed of *wide* graphene nanoribbon (GNR) sections that are *semimetallic*, while the channel regions consist of *narrow* GNR sections that open *semiconducting* bandgaps. Our simulation employs a fully atomistic model of the device, contact and interfacial regions using tight-binding theory. The electronic structures are coupled with a self-consistent three-dimensional Poisson's equation to capture the nontrivial contact electrostatics, along with a quantum kinetic formulation of transport based on non-equilibrium Green's functions (NEGF). Although we only consider a specific device geometry, our results establish several general performance advantages of such monolithic devices (besides those related to fabrication and patterning), namely the *improved electrostatics, suppressed short-channel effects*, and *Ohmic contacts* at the narrow-to-wide interfaces.

## I. INTRODUCTION

IN the past few years, one of the carbon's allotropes, the carbon nanotube (CNT), has created a lot of excitement in the research community as a potential device material replacing or complementing current silicon technology [1,2,3]. CNTs used as interconnect exhibit excellent intrinsic performance, high carrier mobility, high reliability, while CNT field-effect transistors (CNTFETs) exhibit high gain and can be considered a novel device [4]. However, CNTs have yet to impact modern-day electronics because of potentially fundamental difficulties in controlling their *chirality* and *alignment*, leading to complex circuit integration problems. The few experiments that used in-situ growth of CNTs [5,6],

Manuscript received January 25, 2010. The authors acknowledge the support of the Interconnect Focus Center (IFC), one of five research centers funded under the Focus Center Research Program, a DARPA and Semiconductor Research Corporation program and UVa FEST and NSF CAREER grant.

Mircea R. Stan, Avik W. Ghosh, Dincer Unluer, and Frank Tseng are with the Charles L. Brown ECE Department, University of Virginia, Charlottesville, VA 22904 USA (phone: 615 424 0063; fax: 434 924 8818; email: du7x@virginia.edu).

or fluid-flow alignment [7] still show no applicability to general circuit fabrication. Given the semiconductor industry's significant investment in planar fabrication techniques, solutions compatible with current industry practice are clearly preferable. From that perspective, graphene (monolayer graphite) is better suited to current planar fabrication techniques, and, in the form of graphene nanoribbons (GNRs) can exhibit both semiconducting and metallic properties.

Similar to CNTs, GNRs have near-ballistic transport with high mobility (~25000 $cm^2/Vs$ [8] and ~10000 $cm^2/Vs$ [9] have been reported). While all-graphene devices and circuits have been suggested in the literature [8,10], the performance of wide-narrow-wide (WNW) monolithic GNR structures has not been investigated until now. Past research on GNRs has mainly looked at devices that employ bulk metal electrodes as contacts [10,11], which creates Schottky barriers and significant phase incoherence at the device-contact interface.

In this paper, we look at monolithically patterned WNW all-graphene nanoribbon field-effect transistors (GNRFETs) with optimized gate dimensions. A fully atomistic quantum transport model based on non-equilibrium Green's function (NEGF) formalism with 3D electrostatics is applied to explore performance advantages of these devices. Our results show the advantages of using graphene contacts in reducing the source-drain capacitances due to the 2D arrangement, and avoiding Schottky barriers by providing ohmic interfaces, thus allowing the channel and interface energy states to be better dictated by the gate bias. The performance of our devices is characterized by device delay, $I_{on}/I_{off}$ current ratio, and current saturation in current-voltage (I-V) curves for drain voltage sweep. We investigate in depth only combinations of *armchair* GNR (AGNR) structures, with an analysis of *zigzag* GNR (ZGNR) combinations as future work.

This paper is organized as follows: Sec. II briefly discusses the current state-of-the-art in modeling graphene FETs followed by a description of the device structures we investigated, along with the quantum transport formalism based on NEGF, and 3D electrostatics, followed by a brief description of monolithic GNR structures that could form the building blocks for more complex circuits. Section III discusses results and findings obtained with our atomistic modeling. We summarize the main points of our results and provide avenues for future work in Section IV.

## II. MODELING GNRFETS

GNR devices with different kinds of contacts have been studied and implemented by many device physicists and circuit designers. Bulk metallic contacts with all-graphene channel and metallic top and bottom gate have been studied by Guo et al. [10]. GNR metal-semiconductor junctions with different junction geometries were investigated by Guo et al. [12]. The effects of doped graphene contacts on the channel have been studied by Nikonov et al. [13]. An important difference between these previously proposed device structures and ours is that we are using contacts that are undoped and made out of wide GNR regions. Also we examine the effect of these graphene contacts on the narrow graphene channel and derive the I-V characteristics of the GNRFETs. Iannaccone et al. proposed the bilayer graphene tunnel FETs as a device material and obtained large Ion/Ioff ratios with ultralow supply voltages [14]. Experiments indicate that wide GNRs are all *metallic* while ultra thin ribbons (<10 nm in width) are all *semiconducting* and the bandgap increases as the width get smaller [15]; hence a monolithically patterned wide-narrow-wide structure can function as a transistor without the need for atomistic control of the chirality. Experiments by Avouris showed how these nanoribbon devices can be fabricated [16]. De Heer also fabricated arrays of large number of epitaxial graphene transistors on SiC substrate [17]. Russo showed that by using single graphene transistor, operation of four basic two-input logic gates can be achieved [18]. Graphene is a very exciting area of research - recent articles by Barth and Marx [19] and Guo [20] provide an almost exhaustive overview of the breadth and depth of graphene-centered activities and we defer to those papers for an in-depth unified view of the state of the art and for technology exploration in graphene research.

### A. Device Geometry: Wide-Narrow-Wide

We chose to simulate devices patterned *monolithically* from a sheet of graphene with a wide dilution of widths from the source and drain contacts to the active channel region. A metallic gate approximately three times wider than the channel is placed 1nm on top of the channel region, while a wide, but finite, grounded substrate is placed 3nm at the bottom of the channel region to control the device I-V characteristics. Figure 1 shows the device geometry, while the individual band diagrams above each AGNR segment illustrate their respective electronic properties (metallic for the wide regions and semiconducting for the narrow regions) [21].

The metallic and semiconducting electronic properties of GNRs come from boundary conditions on the dominant delocalized electrons at the Fermi wavelength that span four carbon sites [22], created by the interactions of a single $pz$-orbital at each carbon atom. Our main focus is on WNW (35-7-35) GNRFETs composed of (7,0) AGNR *narrow* regions for the channel and slightly *wider*, but still relatively narrow, (35,0) AGNR regions for the contact and interconnect regions. Although a simple tight binding approximation predicts the (35,0) AGNR regions to be semimetallic [22], more detailed modeling and recent experiments [15] show that wider ribbons are needed for true metallicity. The semi-empirical non-orthogonal Extended Huckel Theory (EHT) captures quantitative details of the bandgap that tight-binding theories do not capture, and commonly used *ab-initio* approaches such as Density Functional Theory in the Local Density Approximations (LDA-DFT) underestimate [23]. We have used EHT to model a (35,0) AGNR region and found that it actually exhibits a small $13.47 meV$ bandgap, in disagreement with the single orbital tight-binding, but consistent with experiments [15]. For practical applications we would thus require the contacts and interconnect for GNR circuits to be composed of even wider regions, with strictly metallic bandstructures, but for the purpose of this paper, and for reducing the computation overhead, we revert to simpler tight-binding models that artificially eliminates these small bandgaps to enforce metallicity (in any case, a bandgap less than the room temperature thermal energy of 25$meV$ promotes metallicity). The channel length is set at 8.66$nm$, having 20 unit cells of the (7,0) AGNR. and the width of the (7,0) region is 0.74$nm$. Hydrogen atoms along the top and bottom armchair edges of the device *passivate* the dangling zigzag edge sigma bonds to remove contributions from spurious electron states; such states become evident in more chemically rich EHT model with the physical inclusion of passivating atoms at the edges [22]. In the EHT approach, we also find that the unpassivated pi-bonds lead to an overall 3.5% compressive strain as the GNR edges pick up a benzene-like character (the GNR C-C bond length decreasing from 1.42 Å in bulk graphene to 1.37 Å at the edges, slightly smaller than benzene at 1.39 Å but larger than C=C double bonds at 1.34 Å. (The edges pick up more of a double bond character than benzene because the edges do not enjoy the ring-like resonant symmetry that benzene does) [22]. We also note that metal-induced gap states (MIGS) will still be present because of the contact-channel interfaces, not to be confused with spurious edge states from the broken sigma bonds which are passivated by hydrogen atoms (see Fig. 1, [22]). Despite the presence of MIGS in our WNW all-graphene structure, a semiconducting channel length of 8.66 nm filters those quickly decaying states, resulting in no significant contribution to electron transmission, as seen in Figure 2. Table I summarizes the details of the model structure.

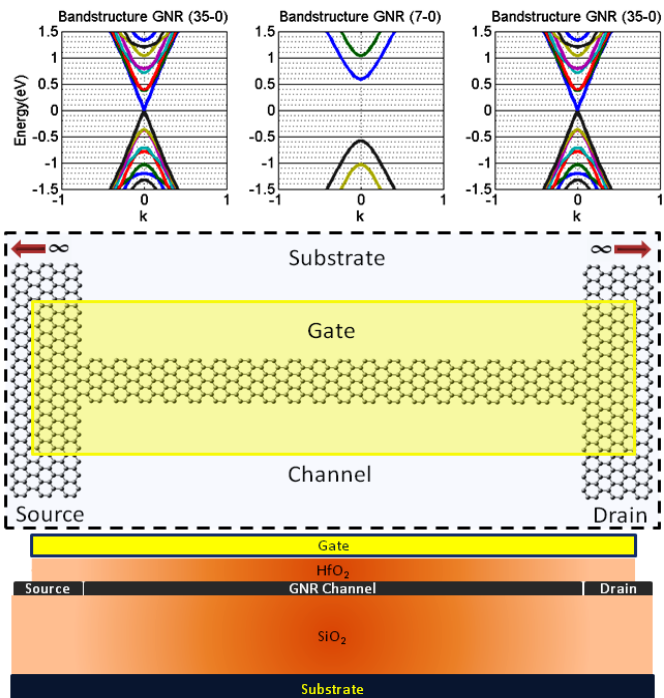

Fig. 1. Monolithic graphene device and interconnect patterned from a single graphene sheet. The gate overlaps the contact regions to achieve better electrostatics. The device region is a semiconducting AGNR sheet and the contacts and interconnect are metallic AGNR segments. The respective tight-binding band dispersions (E-k) are plotted above.

TABLE I
DEVICE DIMENSIONS AND PARAMETERS

| | | |
|---|---|---|
| Channel | Structure | (7,0) Armchair |
| | Width | 0.74 *nm* |
| | Length | 8.66 *nm* |
| Gate | Width | 2.46 *nm* |
| | Length | 10.51 *nm* |
| Source/Drain | Structure | (35,0) Armchair |
| | Width | 4.18 *nm* |
| Gate Insulator | Material | HfO$_2$ (k=16) |
| | Thickness | 1 *nm* |
| Substrate | Material | SiO$_2$ (k=3.9) |
| | Thickness | 3 *nm* |

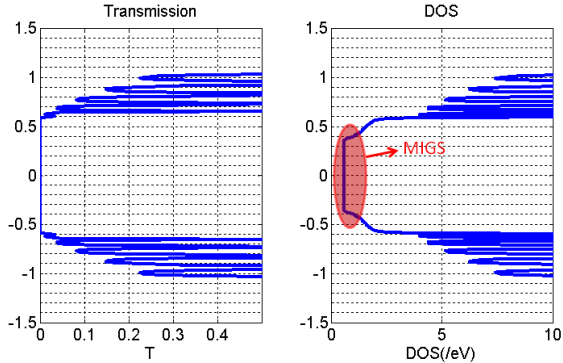

Fig. 2. Showing MIGs that arise from the contact-channel interface, and that are not affected by the hydrogen passivation. MIGs can only be seen in the DOS and not in the transmission.

### B. Quantum Transport: Method for ballistic transport

The approach we use in simulating electron transport through a GNRFET combines the atomistic channel and the contact band dispersion relations with three-dimensional electrostatics and quantum transport within the NEGF formalism [24]. This approach is a significant departure from traditional transport models based on continuum bandstructure (effective mass) and classical drift-diffusion equations that are invalid at nanometer lengths. The method is more applicable for devices with near-ballistic transport, such as the GNRFET which is expected to show long mean-free path due to their band-limited scattering. The atomistic channel is described by a Hamiltonian matrix [H] which accounts for one $pz$-orbital per atom with 3$eV$ coupling between nearest neighbor carbon atoms in a tight-binding approximation (we postpone the use of the more accurate but computationally expensive EHT for future work that focuses on structural issues such as roughness, relaxation, and crumpling in GNRs). Eigenvalues obtained from the Hamiltonian represent the discrete energy levels seen in the band dispersion relations in Figure 1. With a representation of the device energy levels, we define non-Hermitian, energy-dependent self-energy matrices [$\Sigma_{1,2}(E)$] that describe the broadening and shift of the GNR channel energy levels due to coupling with source and drain contacts [24,25]. From the Hamiltonian and self-energy matrices, we compute the device response with the energy dependent retarded Green's Function

$$G = [\, E\,S - H - U_{scf} - \Sigma_1 - \Sigma_2 \,]^{-1} \qquad (1)$$

where $U_{scf}$ is a potential matrix while S is the overlap matrix created by the device basis sets in which case the overlap matrix is  an identity matrix due to fact that tight-binding considers only the $pz$-orbitals. Equation (1) is applied to an NEGF equation for the I-V that integrates the quantum mechanical transmission $T = \text{Trace}(\Gamma_1 G \Gamma_2 G^+)$ [24]

$$I = \frac{2q}{h} \int dE\; T(E)\; (f_1(E) - f_2(E)) \qquad (2)$$

where $f_{1,2}(E)$ are the bias-separated Fermi-Dirac distributions of the contact electrons, and $\Gamma_{1,2} = i(\Sigma_{1,2} - \Sigma_{1,2}^+)$ are the broadening matrices for the channel states. Since the modeled GNRFET has a short channel, we ignore incoherent scatterings in our simulation. The self-energies $\Sigma_{1,2}$ are calculated atomically by solving a matrix recursive equation for the contact surface Green's functions [24,26].

### C. 3D Electrostatics

The potential inside a FET is determined by bias voltages applied to contacts, gate and substrate that drive the device out of equilibrium. To represent the potential inside the channel we compute the Laplace potential and Poisson potential

$$U_L = \frac{c_g}{c_E} V_g + \frac{c_s}{c_E} V_s + \frac{c_d}{c_E} V_d \qquad U_P = \frac{q^2}{c_E} \Delta N \qquad (3)$$

where $C_E$ is the equivalent capacitance of the four parallel electrode capacitances (index g: gate, s: source, d: drain, b: substrate, $\Delta N$: change in the electron number). The Laplace potential ($U_L$) weighs the relative capacitive contributions to the applied bias from the individual electrodes. The Poisson potential ($U_P$) captures the change in the electron density in the channel.  To simplify the computation for atomistic systems, the 3-D Poisson potential is calculated exactly for  a smaller (11-7-11) system that also has a shorter channel length of 2.4 nm. The results are then extrapolated to the larger system at 8.66 nm by using its own density of states to estimate its quantum capacitance, and using the top of the barrier model to scale between system sizes [27]. The quantum capacitance of the smaller system is calculated numerically and then, the large system's quantum capacitance is scaled accordingly to the area change of the large system's channel. The channel potential is solved using the Method of Moments (MOM) by setting up grid points on the channel atoms with a specified charge density, and on the electrode atoms with a specified applied voltage [28]. Since there are two different dielectric regions (one located between the channel and the substrate and other one between channel and the gate), two different equations are needed to capture the contributions of all source charges and their images [29]. Equation (4) below is used when the source charge and the observation points are in the same dielectric material

$$\Phi(r_1, r_2) = \frac{q}{4\pi\varepsilon_0\varepsilon_1}\left[\frac{1}{|r_1 - r_2|} - \frac{\varepsilon_2 - \varepsilon_1}{\varepsilon_2 + \varepsilon_1}\frac{1}{|r_1 - r_2'|}\right] \quad (4)$$

and equation below (5) is used when the source is located in a different dielectric material than the observation points.

$$\Phi(r_1, r_2) = \frac{q}{2\pi\varepsilon_0(\varepsilon_2 + \varepsilon_1)}\left[\frac{1}{|r_1 - r_2|}\right] \quad (5)$$

where $r_2'$ is the distance of the mirror images in the second dielectric from the observation point, assuming both are at the same distance from the boundary. Computed potentials are introduced as diagonal entries into the potential matrix $U_{scf}(1)$.

## III. SIMULATION RESULTS

### A. Optimal Gate Size and High-k dielectric

We begin the analysis of our GNRFET structures by examining the Laplace potential profile across the channel for different gate overlaps. Figure 3 shows the effect of the length of the gate over the potential of the channel. As the gate overlap increases, the local potential spreads out over a longer distance, thus providing better gate control over the channel. Listed in Table I, we used HfO$_2$ (k=16) as a high-k top-gate dielectric and SiO$_2$ (k=3.9) as the substrate dielectric with grounded substrate contact.

### B. Potential Profile: Absence of Schottky barrier

The channel Laplace potential of our GNRFET can be influenced by increasing the source-drain voltage (V$_{ds}$) as well as the gate voltage (V$_g$) as seen in Figure 4. The lowering of the potential throughout the entire channel region with applied gate bias is a characteristic of regular FETs rather than Schottky barrier FETs, whose potentials would otherwise be pinned to midgap by the charging of interfacial states [30]. The ideal C-C bonds at the contact-channel interface in our device create a structure with no pinning states or interfacial de-coherence, *leading to an Ohmic instead of a Schottky contact* (Figure 4).

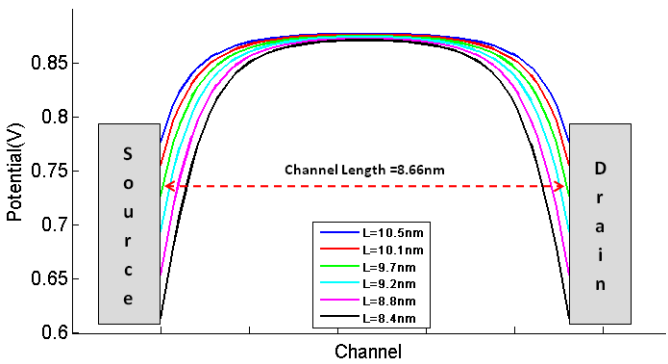

Fig. 3. The different gate overlap sizes with the contacts and their effect on the 8.66$nm$ long channel (narrow) region with V$_g$=1.0$V$. As the gate begins to overlap the contact (wide) regions, the Laplace potential start to smoothen out towards the contacts, thus providing better gate control over the channel.

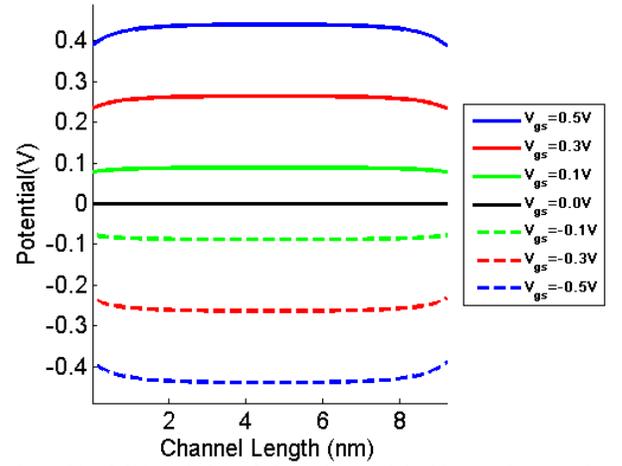

Fig. 4. At V$_{ds}$=0.0V, variation of channel potential with gate shows no barrier pinning at the contacts, implying Ohmic rather than Schottky contacts.

Since the contacts of our monolithically patterned GNRFET are two-dimensional, the charges on the contact surface are line charges, so that the applied source-drain field *decays* into the channel and the corresponding potential is non-linear even in the absence of a gate (Figure 5). This reduces the source-drain capacitance, promoting greater gate control of the channel potential as seen by the flat channel potential with improved short-channel effects. As seen in the Figure 5, the high-k lowers the effects of the source and the drain on the channel thus allowing the gate to be more dominant.

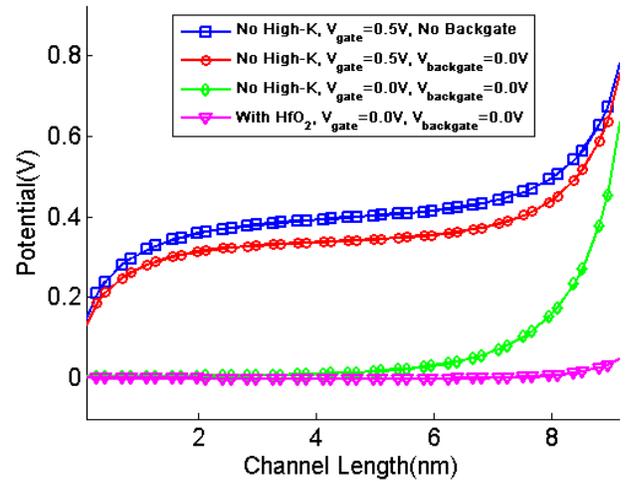

Fig. 5. The two-terminal potential shows vanishing fields near the channel, implying superior gate control and improved short-effects with Vds=1.0V and a topgate length of 10.5nm. Also showing the effects of the backgate and the high-k dielectric on the channel region.

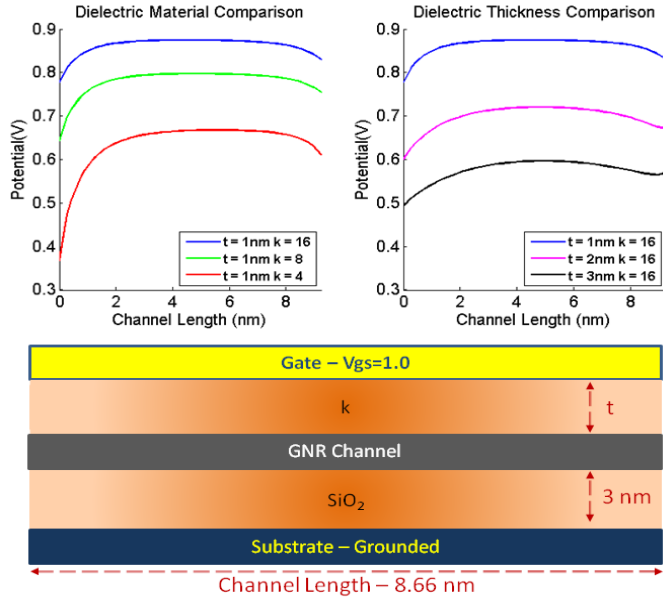

Fig. 6. Effects of different gate dielectric materials and their thicknesses on the channel Laplace potential at $V_{ds}=1.0V$ and $V_{gs}=1.0V$. As the high-k decreases or the thickness of the high-k increases the Laplace potential in the channel is reduced due to the gate losing control over the channel.

The channel Laplace potential can be influenced by the thickness (t) and the high-k (k) of the gate dielectric material. As seen in the Figure 6, as the high-k of the dielectric material is lowered, the gate starts to exercise less control over the channel region thus lowering the potential across the channel. Increasing the thickness of the dielectric material also influences the gate's control over the channel and the effect of the source/drain on the channel. As the thickness is increased the gate loses control and the source/drain starts to control the channel more.

### C. Capacitance: Better gate control

The increased gate control over the potential barrier across the GNRFET channel also indicates better short channel effects. We demonstrate with plots of the channel density of states (DOS) for two scenarios: constant drain voltage ($V_d$) while sweeping gate voltage ($V_g$) in Figure 7a and constant $V_g$ while sweeping $V_d$ in Figure 7b. The sweeping biases of the different contacts create different energy shift rates in the transmission of the GNRFET channel. As expected, sweeping $V_g$ shows larger shifts in the transmission compared to sweeping $V_d$. *This is an outstanding side-effect of the 2-dimensional contacts implicit in the GNR device* and a clear advantage for aggressively scaled technologies.

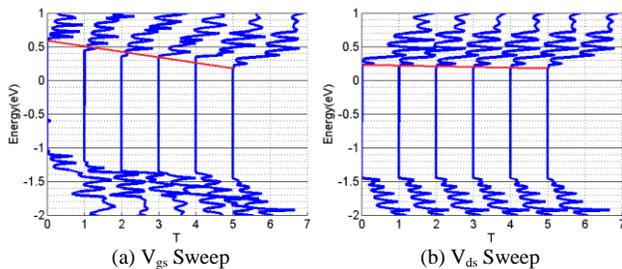

(a) $V_{gs}$ Sweep      (b) $V_{ds}$ Sweep

Fig. 7. An advantage of the 2D electrostatics as seen in (a) is that the gate potential has larger influence on the levels in the channel compared to (b) the applied source/drain potential; thus, demonstrating reduced channeling slipping. Shift in DOS in the channel for voltage sweep used to extract the gate capacitance of $C_g = 3.23 \times 10^{-18} F$ and drain capacitance of $C_d = 9.7 \times 10^{-20} F$. The metallic contacts create non-zero tunneling states in the gap.

From the simulation results showed in Figures 7a and 7b and the charge calculations from the MOM, the $C_g/C_d$ ratio was approximately 33.25, where the same capacitances from (3) represent the relative strengths of the different contacts. Also for the gate control parameter of our model device, we calculated $\alpha_G = 0.94$ which is better than the $\alpha_G = 0.87$ in Rahman's calculations [31] and the $\alpha_G = 0.88$ in Javey's results [27]. For calibration, we compared our results with different, more traditional 3D contact geometry, whose surfaces act as parallel capacitor plates flanked by the insulator at the top and bottom. When such 3D contacts were used, our $C_g/C_d$ ratio dropped to only 24.67 for the same device, gate, and dielectric geometry, proving that 2D contacts indeed help the gate exercise superior control over the channel. The gate control is further improved significantly by optimizing relevant geometric and material parameters, specifically, by increasing its overlap with the contacts (Figure 3) and by using high-k dielectrics.

### D. I-V Curves and Performance: Better short channel effects

The I-V characteristics are calculated using NEGF (Eq. 2), which integrates the quantum transmission function between the source and drain electrochemical potentials [24,25]. Figure 8 shows results for an n-type operation of a smaller (11-7-11) GNRFET system that shows the effects of Poisson on the IVs. As seen in the figure, when the Poisson potential is included the current decreases, but the saturation characteristics improve. Also the Poisson contributions to the IV curves of the small system are non-monotonic with gate voltage. This is because sharp peaks in the DOS arising from contact MIGS cause the $\Delta N$ to not scale linearly with increasing $V_{gs}$. From the difference in the I-Vs, the quantum capacitance of the smaller system and its Poisson contribution is calculated. The Poisson of the smaller system does not contribute much because of the quantum capacitance having a less significant effect on the channel than the gate capacitance. This is due to low DOS in the quantum confined GNR channel [32]. The quantum capacitance is then extrapolated to the larger system with its own density of states by using the top of the barrier model [27]. The scaling approach was adopted to estimate the Poisson contribution for the larger system while avoiding the significant computational burden of doing this atomistically. Our calculated gate and quantum capacitance data agree with the Ref. [32] for comparable geometry.

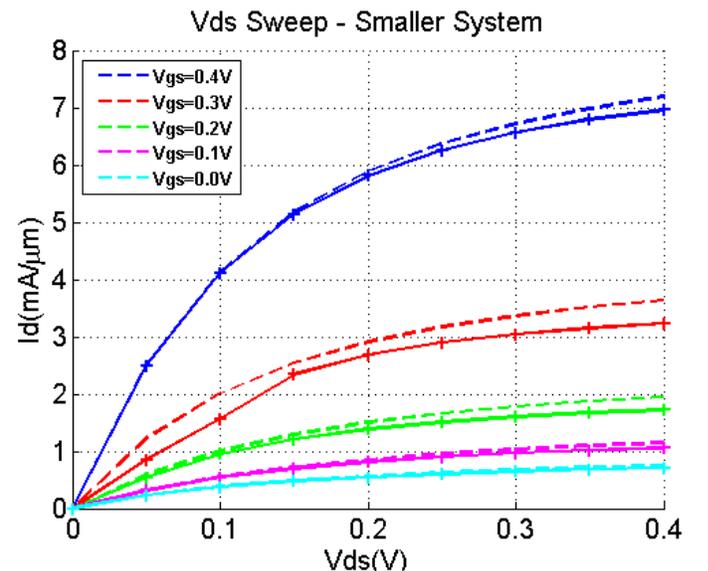

Fig. 8. The Vds sweep for n-type smaller system with the dashed lines presenting only Laplace potential and solid line presenting with the Poisson potential included.

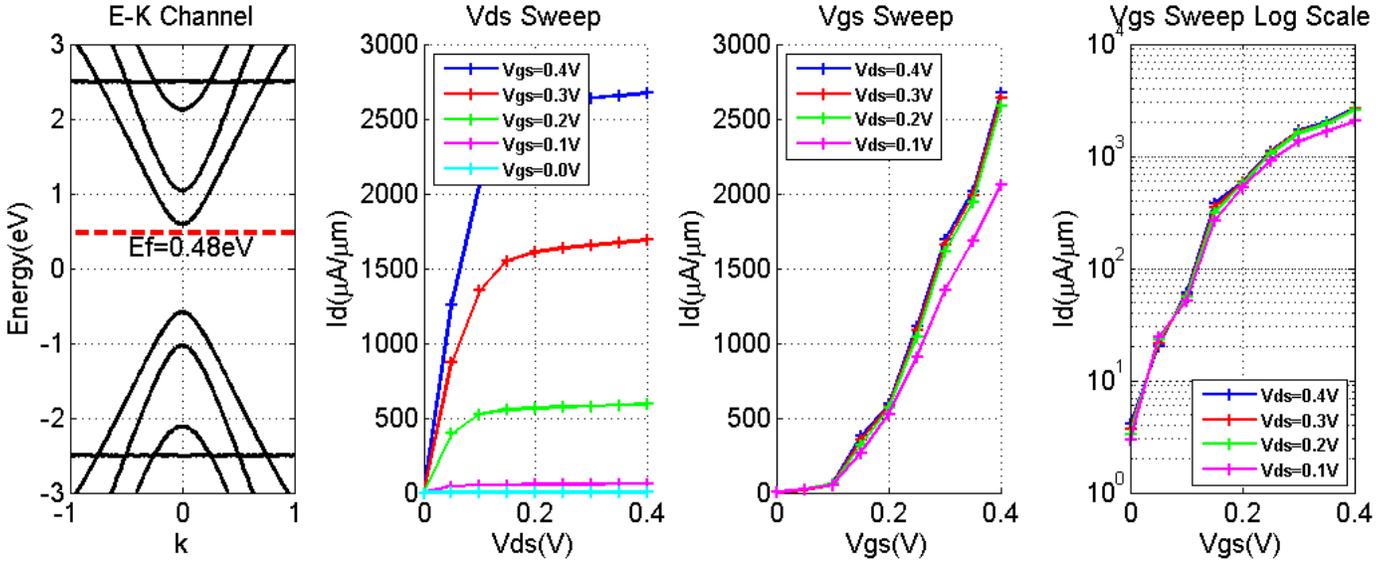

Fig. 9. Band diagram showing the Fermi level and the I-V curves for n-type GNRFET confirming low DIBL and high saturation implying better electrostatics.

While working on the smaller system, it was observed that there were occasional Negative Differential Resistances (NDR). These NDRs arose primarily due to the sharp MIGS states slipping past each other under bias. Since the MIGS states decay rapidly away from the contacts, such NDRs were not encountered in longer ribbons.

Figure 9 shows results for n-type operation of the GNRFET with current saturation and sub-threshold swing (SS) of 84.3$mV$/decade. *Also it should be noted that these devices illustrate ambipolarity of the GNRFET enabled by the lack of Fermi-level pinning.* In our model, n-type and p-type operations are achieved by manually shifting the Fermi-level originally centered between valence and conduction band, by +0.48$eV$ *and* -0.48$eV$ respectively to the edges of the valance and the conduction bands. Traditional semiconductors are doped to shift the Fermi level, which may still be a possibility with adsorbates on GNRs. In the absence of doping, however, this shift can be realized through *electrostatic* doping or by using gate materials with different workfunctions relative to the graphene channel: positive workfunction for n-type operation and negative workfunction for p-type operation. Due to the tight-binding approximation the I-V characteristics of the n-type and p-type GNRFETs simulated in this work are exactly symmetric, but for more accurate models that would no longer be the case.

Simulation results show that the Drain-Induced Barrier Lowering (DIBL) is ~54$mV/V$ which could be further improved by increasing the length of the channel (currently 1:8.6 ratio of HfO$_2$ thickness to channel length). These values are better (smaller) than the estimated values of DIBL=122$mV/V$ and SS = 90$mV$/decade for the double gate, 10$nm$ scaled Si MOSFETs [33]. In addition to the well controlled short-channel effects, the chosen GNRFET structure demonstrates controlled switching between the *on* state and *off* state, with an I$_{on}$ of ~2670.62$\mu A/\mu m$ and I$_{off}$ of ~4.07$\mu A/\mu m$, for an I$_{on}$/I$_{off}$ ratio of ~656. Note that ultrathin GNRs have significant strain at their edges which would increase the band-gap significantly according to EHT predictions [22], further improving the ON-OFF ratio up to 10$^7$. We plan an analysis of the role of structural effects (strain, roughness) on these devices as future work [22].

Another metric for the switching performance of our GNRFETs is the device intrinsic switching delay that can be approximated first-order as C$_g$V$_d$/I$_{on}$. With V$_d$ set at 0.4$V$, and the above I$_{on}$, the GNRFET has an intrinsic device delay of ~0.656$ps$, which is better than current Si-nMOS devices delay at ~0.87$ps$ [10]. The ballistic transit time in the channel is L/v$_F$ ~0.866$fs$, where L=8.66$nm$ is the channel length, and v$_F$ = 10$^8$$cm/s$ is the graphene Fermi velocity. The remaining delay arises during the interfacial injection and removal at the wide to narrow interfaces. Listed in Table II, the dynamic power, $\alpha$C$_g$V$_{dd}^2$f/2, amounts to ~1.25$\mu W$ and the static power, I$_{off}$V$_{dd}$, is ~0.788$\mu W$. The signal to noise ratio of the device should also be quite high due to the high I$_{on}$/I$_{off}$.

An advantage of the device geometry used in this paper (Figure 1) is the presence of covalent carbon-carbon bonds that make up the channel-contact interface. Assuming the highly reactive edge bonds can all be passivated, this configuration avoids Schottky barriers and pinning states typical in metal-semiconductor junctions as seen in Figure 4 [34,35]. The planar patterning of the contacts and channel from a single graphene sheet would instead provide Ohmic contacts, endowing the gate with more control over the channel and interface states. An advantage of the device geometry used in this paper (Figure 1) is the presence of covalent carbon-carbon bonds that make up the channel-contact interface. Assuming the highly reactive edge bends can all be passivated, this configuration avoids Schottky barriers and pinning states typical in metal-semiconductor junctions as seen in Figure 4 [34,35]. The planar patterning of the contacts and channel from a single graphene sheet would instead provide Ohmic contacts, endowing the gate with more control over the channel and interface states. 2D contacts, however, have few modes which can lead to larger voltage drops on them. To accommodate for this loss in the contacts, we need to treat interconnects as series resistances, which will be larger than their 3D counterparts because of the dilution of modes with decreasing dimensionality. In addition, the I-V curves need to be recalibrated to include this series resistance given by $R = \rho_{2d} L/W$, where L is the interconnect length, W is its width, and the 2D sheet resistivity is given by $\rho_{2d} = 1/e\mu n_{2d}$. The 2D mobile electron density $n_{2d}$ in graphite is approximately $10^{11}$-$10^{12}/cm^2$ [8], while the room temperature mobility $\mu$ is as high as ~25,000$cm^2/Vs$ [8]. This results in a sheet resistance of ~2.5(L/W)$k\Omega$. For an interconnect with L/W = 10 in series with the device at $1V$ local voltage (channel current ~$10A$), this results in an additional voltage drop of ~$0.25V$ in the contact, increasing the delay by a factor of ~1.25. Table II summarizes the performance characteristics of the GNRFET structure considered in this paper.

TABLE II
EXTRACTED DEVICE PARAMETERS

| Parameter | GNRFET | Dual-Gate MOSFET [28] |
|---|---|---|
| Ion | 2670.62 μA/μm | 930 μA/μm |
| Ioff | 4.07 μA/μm | 4 μA/μm |
| Ion/Ioff | 656 | 232.5 |
| DIBL | 54 mV/V | 122 mV/V |
| SS | 84.3 mV/dec | 90 mV/dec |
| Device Delay | 0.484 ps/μm | -- |
| Transit Time | 0.866 fs | -- |
| Dynamic Power | 1.25 μW | -- |
| Static Power | 1.628 μW/μm | 1.628 μW/μm |

## IV. SUMMARY AND FUTURE WORK

In this paper we conducted a study of a GNRFET structure patterned monolithically out of a single sheet of graphene with a metallic top gate over the channel region and a back gated substrate. To conduct this study we applied a full quantum coherent transport model using NEGF and 3D electrostatics for a specific atomic structure and device geometry.

Since wide GNRs exhibit 3-fold periodic symmetry for metallicity we were able to use metallic GNRs for the source-drain contacts, while confinement and strain open bandgaps in narrow ribbons [22], which allows us to choose a semiconducting GNR for the channel. A particular (35-7-35) AGNR structure was chosen as an example, but the goal is to move towards generic wide-narrow-wide GNR structures that lead to similar characteristics without the need for atomistic control of the width of the ribbon. The carbon-carbon interfacial bonds at the channel-contact interface implicitly avoid phase-breaking processes associated with Schottky barriers; instead we have Ohmic contacts modeled as series resistances. Various simulations of the potential profile and density of states with sweeping voltages show the increased gate control and decreased influence from the two dimensional source-drain contacts. Device performance metrics such as device delay and $I_{on}/I_{off}$ ratio reinforce the superior switching ability of this type of GNRFET. The improved DIBL is due to the better gate control compared to the drain on the channel. In contrast to carbon nanotubes, the switching between metallic and semiconducting behavior simply needs a wide modulation of ribbon widths without the need for atomistic control of the edge state geometry. In addition, the ability to pattern these ribbons with a combination of top-down lithography and bottom-up edge-state chemistry in principle allows the fabrication of circuits with adequate on currents for fast switching, circumventing Schottky barriers in CNTFETs. Future work will look at optimizing our monolithically patterned GNRFET geometry, analyzing more complex AGNR and ZGNR structures, studying the influence of structural anomalies such as edge roughness, passivation, strain, and crumpling, exploring the use of electrostatic doping and gate workfunction choice for determining n-type and p-type behavior, and utilizing extracted compact model parameters to design and optimize circuit level performance metrics of all-graphene circuits, where possibly even the gate regions are fabricated out of wide, semi-metallic graphene sheets.